\documentstyle[leqno,twoside,11pt,ltexproc]{article} 

\begin{document}
\cleardoublepage
\pagestyle{myheadings}

\title{Classical and Quantum Properties of Liouville Black Holes
\thanks{work supported by
the Natural Sciences and Engineering Research Council of Canada}}
\author{R.B. Mann\thanks{Physics Dept., University of Waterloo,
Waterloo, Ontario, Canada N2L 3G1}}
\date{WATPHYS-TH-94/03}
\maketitle
\markboth{Mann}{Properties of Liouville Black Holes}
\pagenumbering{arabic}

\begin{abstract}
Black hole spacetimes can arise when a Liouville field  is coupled to two-
dimensional gravity. Exact solutions are obtained both classically and when
quantum corrections due to back reaction effects are included. The
black hole temperature depends upon the mass and the
thermodynamic limit breaks down before evaporation of the
black hole is complete, indicating that higher-loop effects must be
included for a full description of the process.
\end{abstract}

Liouville field theory  has been a useful tool in expanding our
understanding of 2D quantum gravity. The usual approach is to consider the
Liouville field to be the conformal factor of the metric, whose quantum
properties are then derived from the quantum dynamics of the Liouville
field coupled to other 2D matter \cite{Polya,DDK}.

Recently a different approach has been adopted in which  the Liouville
field is taken be an independent matter field whose stress-energy couples
to 2D gravity in a manner somewhat analogous to the $(3+1)$-dimensional
case \cite{RtDil,lblack}.  Specifically, the classical field equations
of a Liouville field in curved spacetime are modified to include quantum
corrections due to both conformally coupled matter and to the
gravity/Liouville system itself. The field equations contain exact
solutions in each case which correspond to asymptotically flat black
holes whose temperature depends upon their ADM-mass.

The action is a particular version of dilaton gravity \cite{MST,semi}
in which
\begin{equation}
S = S_G + S_M = \frac{1}{8\pi G}\int d^2x\sqrt{-g}\left(
\frac{1}{2} g^{\mu\nu}\nabla_\mu\psi
   \nabla_\nu\psi +\psi R \right) +S_M\label{1}
\end{equation}
where $R$ is the Ricci scalar and the matter action $S_M$ is independent of
the auxiliary dilaton field $\psi$. After some manipulation the field
equations are easily seen to be
\begin{equation}
R = 2\pi T_\mu^{\ \mu}  \qquad \qquad \nabla_\nu T^{\mu\nu} = 0 \label{2}
\end{equation}
\begin{equation}
\frac{1}{2}\left(\nabla_\mu\psi\nabla_\nu\psi - \frac{1}{2} g_{\mu\nu}
(\nabla\psi)^2\right) + \frac{1}{2}
g_{\mu\nu}\nabla^2\psi - \nabla_\mu\nabla_\nu\psi
= 2\pi (T_{\mu\nu}-\frac{1}{2}g_{\mu\nu}T^\lambda_{\lambda}) \label{3}
\end{equation}
where $T_{\mu\nu}$ is the conserved 2D stress-energy tensor associated with
$S_M$. Note that the evolution of the gravity/matter system is independent
of $\psi$.

Taking the matter action to be that of a Liouville field $\phi$,
\begin{equation}
S_M = S_L = \int d^2x \sqrt{-g}[b (\nabla\phi)^2
+ \Lambda e^{-2a\phi} + \gamma \phi R] \label{1a}
\end{equation}
(which reduces to the standard Liouville action for 2D flat space),
the field equations (\ref{2},\ref{3}) become (after some manipulation)
\begin{equation}
(b + 4\pi G\gamma^2)\nabla^2\phi = (4\pi G\gamma -a)\Lambda e^{-2a\phi}
\label{9}
\end{equation}
\begin{equation}
(b + 4\pi G\gamma^2) R = 8\pi G\Lambda (a\gamma + b) e^{-2a\phi} \label{10}
\end{equation}
with the evolution of $\psi$ being determined by the traceless part of
(\ref{3}).

In coordinates where $\partial/\partial t$ is a Killing vector one can
write the metric in the form  $ds^2 = -\alpha(x) dt^2
+ {dx^2}/{\alpha(x)}$. One can obtain exact solutions to the field
equations for all possible values of the parameters $a$, $b$, $\gamma$
and $\Lambda$ (except for a set of measure zero) \cite{lblack}.
These are given in table I below.
\medskip
\bigskip
\centerline{\bf Table I}
\medskip
\noindent
\begin{tabular}{||l|c|c|c||}\hline
&({\bf A})&({\bf B})&({\bf C})\\
\hline\hline
$\alpha(x)$ & $2Mx - Bx_0^2(\frac{x}{x_0})^p $
& $1 - Ce^{-2M(x-x_0)}$ & $2Mx\ln(2D x x_0)$ \\
\hline
$\phi$ & $\frac{2-p}{2a}\ln(x/x_0) + \phi_0$ & $\frac{M}{a}(x-x_0) + \phi_0 $
& $\frac{1}{2a}\ln(\frac{4D}{M}x) $\\
\hline
$\psi$ & $-p \ln(2Mx) +\psi_0$ & $2Mx+\psi_0$ & $-2M\ln(\frac{D}{M}x)
+\psi_0$\\
\hline
Parameters& $p = 8\pi G \frac{a\gamma+b}{a^2+4\pi
Gb} $ & $b=-\frac{a^2}{4\pi G}$ & $b=\frac{a^2}{4\pi G}-2a\gamma$ \\
& $B = \Lambda \frac{(a^2+4\pi G b)^2}
{(b+4\pi G\gamma^2)(4\pi Gb+8\pi Ga\gamma - a^2)}$&
$C \equiv \frac{2\pi G\Lambda a}{M^2(a+ 4\pi G\gamma)}$ &
$D = \pi G\Lambda\frac{a}{4\pi G \gamma - a}$\\
\hline
${\cal M}$&$M\frac{8\pi G\gamma(\pi G{b}+ 4\pi G a\gamma)+ a^3 -10\pi G\gamma
a^2}
{4\pi G a(a^2 + 4\pi G{b})} $ & $\frac{M}{8\pi G}(1- 8\pi G\gamma/a)$
& $\frac{M}{8\pi G}(1- 4\pi G\gamma/a)$ \\
\hline
Temperature &$2\hat{K}{\cal M}$ & $4G{\cal M}\frac{a}{a-8\pi G\gamma}$
  &$4G{\cal M}\frac{a}{a-4\pi G\gamma}$\\
\hline
Entropy& $\frac{1}{2\hat{K}}
\ln({\cal M}/{\cal M}_0)$
& $\frac{a-8\pi G\gamma}{4Ga}\ln({\cal M}/{\cal M}_0)$
& $\frac{a-4\pi G\gamma}{4Ga}\ln({\cal M}/{\cal M}_0)$ \\
\hline
\end{tabular}

\noindent
where $\hat{K}$, $\zeta_1$ and $\zeta_2$ are constants formed from
$a$, $b$, and $\gamma$. One can replace $x$ by $|x|$ in the above solutions
by including an additional point-like source at the origin \cite{lblack}.

The criteria for a solution in
table I to be a black hole solution are as follows.
{\bf (1)} The spacetime must be asymptotically flat, {\it i.e.}
$R$ should vanish as $x\to \infty$.
{\bf (2)} There must be an event horizon for a finite real value of $x$.
{\bf (3)} The metric signature must be $(-,+)$ for large $x$; otherwise the
horizon is a cosmological horizon.
{\bf (4)} The ADM-mass ${\cal M}$ must be real and positive.
The temperature and entropy in table I apply only to the subset of
solutions which are black hole solutions.

In contrast to string-theoretic black holes \cite{MSW,CGHS} (whose
temperature is constant) the temperature of Liouville black holes varies
linearly with their ADM-mass, leading to striking different thermodynamic
behaviour \cite{TomRobb}.
Consider a box of length $L$ containing a Liouville black hole of
mass $M$ and thermal radiation at temperature $T$. Scaling out the
parameter dependence, the energy and entropy of this system are
\begin{equation}
{\cal S}=2\pi \ln \Bigl( {M\over M_0}\Bigr) + {\pi\over 3} TL
{}~~~~~~~E=M+{\pi\over 6} T^2L  \label{eq14}
\end{equation}
Maximizing ${\cal S}$ for fixed $E$ yields
\begin{equation}
0={\partial S\over \partial M}={2\pi\over M} - {1\over T} ~~
\Rightarrow ~~T={M\over 2\pi} \label{eq15}
\end{equation}
recovering the result in table I. Note that
${\partial ^2S\over \partial M^2} =-{2\pi\over M^2}-{3\over \pi LT^3} <0$,
guaranteeing a maximum. Hence entropy is always maximized provided
(\ref{eq15}) can be physically realized. It is also possible to show that a
phase transition between pure radiation and the combined black-hole/radiation
scenarios can also take place \cite{TomRobb}.

The entropy of a Liouville black hole varies logarithmically with its
mass. Hence it is possible for the entropy of 2 widely separated Liouville
black holes to be greater than a single hole of equivalent ADM mass,
corresponding to a qualitatively new type of thermodynamic system. In normal
thermodynamic systems the entropy is a superadditive, homogenous and
concave function of the extensive variables (for black holes, this is
${\cal M}$). For $(3+1)$ dimensional black holes the latter two properties
are violated; however for Liouville black holes (and $(2+1)$-dimensional
black holes) the former two properties are violated (or, under certain
circumstances, only homogeneity  is violated) \cite{ptl}.

Quantum corrections to the black hole evaporation process have been
considered under two distinct scenarios, by incorporating the stress-energy
of $N$ quantized scalar fields or by performing a full (1 loop)
quantization of the gravity/Liouville system (\ref{1},\ref{1a}). In each
case the Newtonian constant $G$ gets renormalized and the space of
solutions in table I maps into itself \cite{lblack}.

The inclusion of the quantum stress-energy typically
results in a larger ADM-mass, which decreases with time as
${\cal M} \sim 1/t$. Eventually the proper distance from the
singularity to the horizon (the ``size'' of the black hole)
becomes comparable to the Compton wavelength of a particle of
similar mass. Taking this to be ${\cal M}_0$ in (\ref{eq14}),
the entropy tends to zero for finite temperature.
At this point the
thermodynamic description breaks down and higher-loop corrections become
important. Previous work \cite{MST} suggests that inclusion of
quantum vacuum energies might cause the black hole to
slowly cool off to a zero temperature remnant, leaving
behind a global event horizon with its requisite loss of quantum coherence.
However the details of such a scenario remain to be investigated.

\end{document}